\documentclass[final,3p,times,twocolumn]{elsarticle}

\usepackage{graphicx}

\usepackage{amssymb,amsfonts,amsmath,listings}
\usepackage{color,times,float}
\usepackage{orcidlink}
\newlength{\figurewidth}
\newlength{\pagewidth}
\definecolor{dkgreen}{rgb}{0,0.6,0}
\definecolor{gray}{rgb}{0.5,0.5,0.5}
\definecolor{mauve}{rgb}{0.58,0,0.82}
\definecolor{backcolour}{rgb}{0.97,0.97,0.97}
\newcounter{bla}

\journal{Computer Physics Communications}

\begin{document}

\begin{frontmatter}

\title{A Software Package for Generating Robust and Accurate Potentials using the Moment Tensor Potential Framework}

\author[buf]{Josiah~Roberts\,\orcidlink{0000-0002-1359-5184}}
\author[buf]{Biswas~Rijal\,\orcidlink{0000-0003-3679-4191}}
\author[mems,camd]{Simon~Divilov\,\orcidlink{0000-0002-4185-6150}}
\author[psu]{Jon-Paul~Maria\,\orcidlink{0000-0003-3604-4761}}
\author[mst]{William~G.~Fahrenholtz\,\orcidlink{0000-0002-8497-0092}}
\author[psu]{Douglas~E.~Wolfe\,\orcidlink{0000-0002-3997-406X}}
\author[ncsu]{Donald~W.~Brenner\,\orcidlink{0009-0009-1618-4469}}
\author[mems,camd]{Stefano~Curtarolo\,\orcidlink{0000-0003-0570-8238}}
\author[buf]{Eva~Zurek,\orcidlink{0000-0003-0738-867X}\corref{author}}


\cortext[author] {Corresponding author.\\\textit{E-mail address:} ezurek@buffalo.edu}
\address[buf]{Department of Chemistry, State University of New York at Buffalo, Buffalo, NY 14260, USA}
\address[mems]{Department of Mechanical Engineering and Materials Science, Duke University, Durham, NC 27708, USA}
\address[camd]{Center for Autonomous Materials Design, Duke University, Durham, NC 27708, USA}
\address[psu]{Department of Materials Science and Engineering, The Pennsylvania State University, University Park, PA 16802, USA}
\address[ncsu]{Department of Materials Science and Engineering, North Carolina State University, Raleigh, NC 27695, USA}
\address[mst]{Department of Materials Science and Engineering, Missouri University of Science and Technology, Rolla, MO 65409, USA}

\begin{abstract}

We present the Plan for Robust and Accurate Potentials (PRAPs), a software package for training and using moment tensor potentials (MTPs) in concert with the Machine Learned Interatomic Potentials (MLIP) software package. PRAPs provides an automated workflow to train MTPs using active learning procedures, and a variety of utilities to ease and improve workflows when utilizing the MLIP software. PRAPs was originally developed in the context of crystal structure prediction, in which one calculates convex hulls and predicts low energy metastable and thermodynamically stable structures, but the potentials PRAPs develops are not limited to such applications. PRAPs produces two potentials, one capable of rough estimates of the energies, forces and stresses of almost any chemical structure in the specified compositional space -- the Robust Potential -- and a second potential intended to provide more accurate descriptions of ground state and metastable structures -- the Accurate Potential. We also present a Python library, \textit{mliputils}, designed to assist users in working with the chemical structural files used by the MLIP package.

\end{abstract}


\end{frontmatter}

{\bf PROGRAM SUMMARY}

\begin{small}
\noindent
{\em Program Title:} The Plan for Robust and Accurate Potentials (PRAPs) \\
{\em CPC Library link to program files:} (to be added by Technical Editor) \\
{\em Developer's repository link:} https://github.com/Dryctarth/PRAPs.git \\
{\em Code Ocean capsule:} (to be added by Technical Editor)\\
{\em Licensing provisions(please choose one):} BSD 3-clause \\
{\em Programming language:} Bash, Python                                   \\
{\em Supplementary material:} User manual                                 \\
{\em Nature of problem:} Keeping track of all the steps involved in training moment tensor potentials across several systems has proven to be a challenge in need of project management. For every large step, like training, there are several small, mundane commands that need to be handled, and these must all be repeated identically across any chemical system users may care about (while tracking variations). Finally, communication must be made between the AFLOW, MLIP, and VASP programs.\\
{\em Solution method:} The PRAPs package incorporates a degree of automation, handling the different job submissions and tasks needed to train multiple moment tensor potentials, file management, identifying and removing unphysical chemical structures, and performing some analytical tasks. The package also includes some simple utility functions to allow users to better read, write, and manipulate MLIP's chemical structure file format.\\
{\em Additional comments including restrictions and unusual features:} Requires a local installation of Automatic FLOW (AFLOW) v3.10+ , the Vienna \emph{ab initio} Software Package (VASP) v5+, and the Machine Learning for Interatomic Potentials (MLIP) v2+ program packages.\\
   \\

\end{small}

\section{\label{sec:Intro}Introduction}
The application of machine learning (ML) techniques towards materials research has steadily gained importance in the last decade~\cite{Curt1,Schmidt:2019a}. Though there are numerous ways in which ML promises to enable and accelerate the discovery of materials with a particular functionality, herein we focus on its use in atomistic calculations. In the past density functional theory (DFT) calculations have been the method of choice in materials research, providing a reasonable balance between accuracy, speed, and scaling. However, they are not sufficiently fast for high-throughput materials discovery of complex multi-component systems, and are limited in the size of the simulation cell and the length-scale that can be employed in molecular dynamics simulations. ML interatomic potentials (ML-IAPs), trained on DFT data, promise to overcome these limitations, while providing advantages to traditional potentials with fixed functional forms~\cite{Zuo,HartNPJ,Mishin:2021a}. {Indeed, numerous high-throughput workflows for materials modelling that support ML-IAPs have recently been released~\cite{li2025apex,ganose2025atomate2,janssen2019pyiron}, some of which include workflows for training~\cite{liu2025automated} and fine-tuning~\cite{li2025apex} ML-IAPs.} 

A zoo of ML-IAPs, which can be trained for a particular system on a user-specified set of DFT data to generate bespoke models have been proposed~\cite{Behler:2007a,Hajinazar2017,Thompson:2015a,Bartok:2010a,xie2023ultra,Pickard2022}. More recently, off-the-shelf ML-IAPs, which can predict energies, forces and stresses (EFS) across the periodic table have come to the fore~\cite{m3gnet,chgnet,alignn,GoogUniverse}. Though these potentials tantalize, as they do not require the generation of DFT data for training, their performance in extrapolating to complex atomic environments, which are out of distribution, is unknown. Therefore, it is unclear if such universal ML-IAPs would be useful in crystal structure prediction (CSP), in particular in extreme environments, such as high pressures. Many of the global search metaheuristics~\cite{Zurek:2014d} such as evolutionary or genetic algorithms \cite{Glass:2006a,SH07,Zurek:2020i,Tipton:2013a}, particle swarm methods~\cite{Wang:2010a}, and random searches~\cite{Pickard:2011a} require the generation of a diverse set of structures, which need to be locally minimized, for the discovery of the (putative) global minimum and intriguing local minima. 

The Moment Tensor Potentials (MTP)~\cite{Shapeev:2016a,MLIP1} flavor of ML-IAPs are sometimes used because of their excellent balance between model accuracy and computational efficiency~\cite{Zuo,HartNPJ}. They have been successfully applied to study multi-component systems~\cite{MTP+HEA, gubaev2023performance,Pandey:2022a,Guo:2023a,Zheng:2023a,Meziere:2025a,Liu:2024a}, including predicting phonons and thermodynamic properties~\cite{TherMTP,MTPhonons,MTP+TILD}, and their use in CSP is noteworthy~\cite{Gubaev:2019a,Podryabinkin:2019a,Zurek:2023n}. The Machine Learned Interatomic Potentials (MLIP) software package~\cite{MLIP2} can be employed to train MTPs on a user-specified DFT training set (either via basic training or active learning), and to perform local relaxations. Moreover, MLIP can be interfaced with other packages, most notably LAMMPS for relaxations and molecular dynamics simulations~\cite{LAMMPS,MLIP2}.

\begin{figure}
    \centering
    \includegraphics[width=0.8\columnwidth]{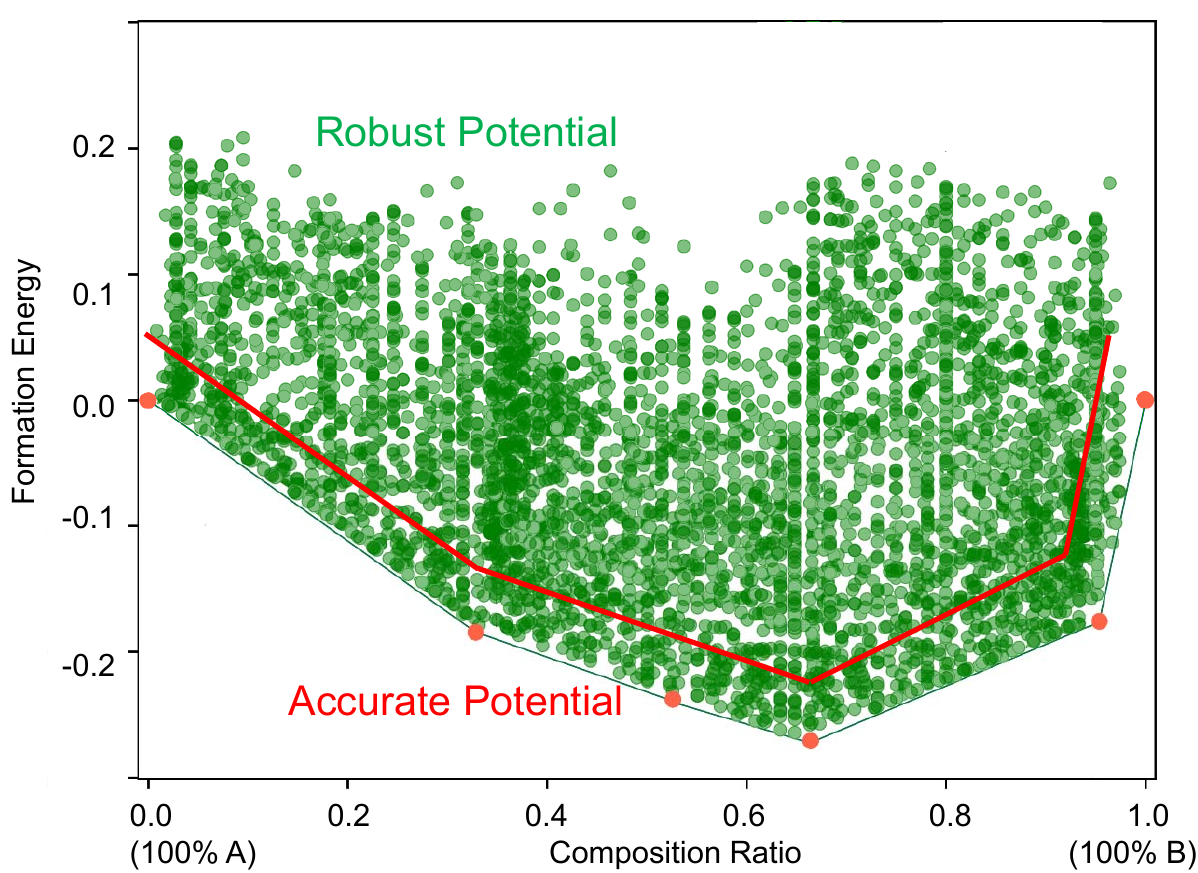}
    \caption{An illustration of the compositional spaces used for training in a two-component system. The robust potential (\texttt{RP}) is trained on all structures. The accurate potential (\texttt{AP}) is trained on structures below the red line: those within a certain energy from the minimal energy structure of that composition. The solid line is the convex hull, and the red dots represent the structures that lie on it.}
    \label{fig:ap}
\end{figure}

CSP using global optimization schemes that traverse the potential energy surface requires the optimization of structures that are both high and low in energy. Generating a ML-IAP that can accurately predict the EFS of such a wide variety of structures is extremely challenging. For this reason a pragmatic approach might consist of  approximately computing the energies of a large number of structures, and using these rough values to determine which should be discarded, and which should be kept for more accurate energy predictions. One specific strategy has been to re-rank the energies of a subset of the ML-IAP most stable structures with DFT, followed by DFT optimizations of yet a smaller subset~\cite{Salzbrenner}. Another proposed strategy has been to generate ML-IAPs that are robust (for the optimization of any configuration, but with a potentially large associated error), followed by those that are accurate (to be used for the more precise prediction of the energies of structures that lie near the convex hull)~\cite{Gubaev:2019a, Podryabinkin:2019a}. A schematic illustration of a two-component convex hull and the energies of the structures or configurations that could be predicted by a robust potential (\texttt{RP}) and an accurate potential (\texttt{AP}) is provided in Figure \ref{fig:ap}.

We have recently written a utility package to be used for the second strategy. This package automates the training of such \texttt{RP}s and \texttt{AP}s, and it is therefore aptly named \emph{The Plan for Robust and Accurate Potentials} (PRAPs). PRAPs was applied towards the prediction of the zero-temperature zero-pressure phase diagrams of four ternary metal carbides (CHfTa, CHfZr, CMoW and CTaTi)~\cite{Zurek:2023n}. While the initial manuscript that employed PRAPs provided prediction errors for both the \texttt{AP} and \texttt{RP} for MTPs of differing complexity (levels) for these carbides, the utility package was not described nor released. Herein, we describe the composition and usage of this utility package, and make it available for the broader scientific community.  PRAPs provides a workflow for training and analysis of \texttt{RP}s and \texttt{AP}s, and handles many of the mundane tasks where fatigue and carelessness are prone to derail human operators. It also interfaces with the Vienna \emph{ab initio} Software package (VASP~\cite{VASP}) and the Automatic Flow package (AFLOW)~\cite{AFLOW,AFLOWstd} for automating DFT calculations. 

\section{MTP Training}
ML-IAPs of the MTP flavor have become one of the methods of choice in materials science applications.  Detailed information about their mathematical construction can be found in original work by Shapeev~\cite{Shapeev:2016a,MLIP2}. The active learning scheme (ALS), which can be used to improve an MTP during the course of a simulation  (structural relaxation or molecular dynamics run), has been described in Reference~\cite{MLIP1}. The purpose of this section is to provide a brief overview of the MTP method, covering only those aspects required to understand the workflow of the PRAPs package. 

The simplest form of MTP training performed by the MLIP software package is basic training. Here, an MTP is trained on a set of chemical structures containing energy, force, and stress (EFS) data, usually obtained from DFT. These structures are stored in a .cfg (short for configurations) file. Since a .cfg file may contain numerous entries for a particular structure, for example when a relaxation trajectory is used, the word `configuration' denotes a particular set of atoms, their coordinates, and any associated EFS data, keeping in mind that numerous configurations might be associated with a single structure.

The number of basis functions, and therefore the number of parameters that need to be determined during the training, grows exponentially with the MTP level, abbreviated as lev$_\text{max}$. This level also affects the number of radial functions used to construct the MTP. As a result, training at higher levels will take longer, but will be able to benefit from larger datasets. During training, the fitting parameters are obtained by minimizing the difference between the EFS data predicted by the MTP and the training data. When MLIP initializes the training process starting from an untrained MTP the fitting parameters are chosen randomly, meaning that the MTPs from two basic trainings may differ. For this reason, it has been suggested that basic training should be repeated five times. Comparison of the errors and predictions from the five trainings can help determine training reliability, and find the MTP with the lowest error~\cite{MLIP2}.

In addition, MLIP can train MTPs using active learning. Active learning is a process in which an MTP is iteratively re-trained across a use-evaluate-train loop. The MTP is used to relax a set of structures. During the relaxation, the MLIP software estimates a degree of extrapolation (or grade), $\gamma$, of every structure that is encountered. This grade is based on the D-optimality criterion~\cite{MLIP1,MLIP2}, which lets MLIP choose those configurations that are neither too similar nor too different from those already in the training set. The grade feature is also employed to determine where the extrapolation is deemed to be risky, terminating the relaxation of an individual structure. While users may set their own grade thresholds, the MLIP authors recommend $\gamma_\text{select}\ge 2$ and $\gamma_\text{break}=10$. Upon the completion of the relaxation command, regardless of whether all structures successfully relaxed or not, configurations with grades between 2 and 10 are chosen and sent for DFT single point calculations. The resulting configurations with new DFT data are added to the initial training set, and the resulting set is used to retrain the MTP. The active learning procedure repeats until the MTP-based simulation completes without the need for retraining. With these points in mind, let us now proceed to describe the PRAPs workflow.

\section{Plan for Robust and Accurate Potentials (PRAPs)}
\subsection{Workflow} \label{sec:workflow}

PRAPs typically begins with a set of atomic configurations for which DFT data is already available, though the procedure can begin with a set of configurations without EFS data. The EFS may have been obtained by the user, or scoured from an online database. For example, previously~\cite{Zurek:2023n} we employed relaxation trajectories for compounds with the C-Hf-Ta, C-Hf-Zr, C-Mo-W, and C-Ta-Ti elemental combinations from the AFLOW database~\cite{AFLOW,curtarolo:art190,curtarolo:art191} to pre-train an MTP. This data will be kept in a .cfg file, used by the MLIP program, and is called \textbf{DFT\_CFG} in PRAPs. 
\begin{figure}
    \centering
    \includegraphics[width=0.8\columnwidth]{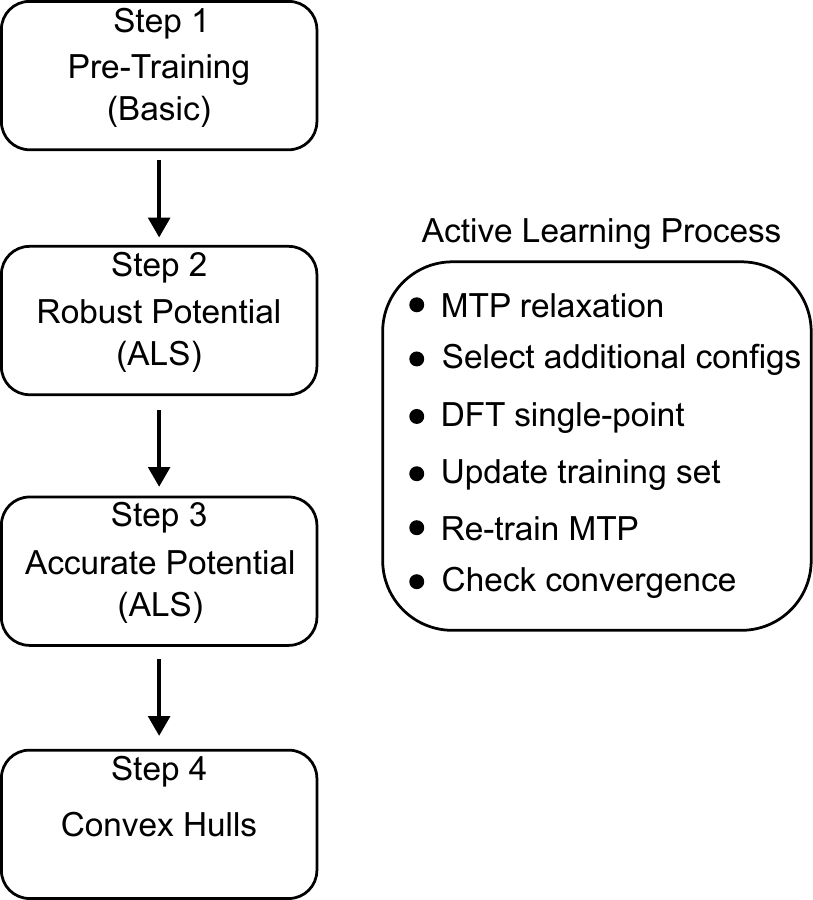}
    \caption{Workflow in the \emph{Plan for Robust and Accurate Potentials} (PRAPs) package, which automates the generation of a moment tensor potential (MTP), given any quantum mechanical training set. Step 1: Five MTPs are trained on a set of configurations and the best (the pre-Robust Potential, \texttt{pre-RP}) is chosen. Step 2: The \texttt{pre-RP} is employed to initialize the training of the Robust Potential (\texttt{RP}) via an active learning scheme (ALS, inset). Step 3: The \texttt{RP}-relaxed lowest energy configurations  are chosen to train an Accurate Potential (\texttt{AP}) via active learning. Step 4: Convex hulls may be generated using the \texttt{AP} and AFLOW.}
    \label{fig:workflow}
\end{figure}

PRAPs operates four main processes, each separated by checkpoints should a job be interrupted partway through. First is \emph{Pre-Training}, in which a set of MTPs are generated via basic training and the best is selected for use in active learning. Second, PRAPs trains the \emph{Robust Potential} (\texttt{RP}) using active learning. Third, PRAPs relaxes all provided configurations with the \texttt{RP} and uses active learning again to train the \emph{Accurate Potential} (\texttt{AP}). Finally, PRAPs uses the \texttt{RP} and \texttt{AP} to perform some analysis, including generation of convex hulls, if desired. A general workflow is provided in Figure 2, while the remainder of this section will describe each of these steps in detail.

\textbf{Step 1:} The first step of the PRAPs procedure is to generate an MTP with basic training. This MTP, the \emph{Pre-Robust Potential} (\texttt{Pre-RP}), will be used as the starting MTP for the active learning steps. PRAPs will train five potentials on data from the \textbf{DFT\_CFG} set and select the best one. To generate the training sets structures are selected randomly from the filtered  \textbf{DFT\_CFG}, with instructions to not select any configurations within four indices of each other (to avoid selecting consecutive steps from a relaxation). The ``best'' MTP is defined as the one that can most correctly rank the ten highest and ten lowest energy configurations. In the case of a tie or ambiguity in the ranking, the MTP with the smallest root-mean-squared training error is used. A user may actually skip this entire step if desired. Data filtration is described fully in Section \ref{sec:filt}.

\textbf{Step 2:} The second step is the generation of the \texttt{RP} via active learning. The active learning scheme is described in more detail elsewhere~\cite{MLIP2}, and is used twice in PRAPs. In general, active learning starts by relaxing a set of configurations with a given MTP. During the relaxation, any configurations whose $\gamma$ falls between 2 and 10 are set aside into the \emph{pre-selected set}. The D-optimality criterion is then employed to select which configurations from this set should be added to the \emph{training set}. EFS data is obtained for compounds comprising the \emph{selected set} by single-point DFT calculations. The training set is updated with these new configurations, the MTP is re-trained, and the cycle continues. PRAPs uses this workflow to train the \texttt{RP} and to train the \texttt{AP}.

The \texttt{RP} training is performed by starting with the \texttt{Pre-RP} MTP, the training set it was trained on (\textbf{DFT\_CFG}), and any additional structures that do not have EFS data from DFT (the unrelaxed set, or \textbf{URX\_CFG}). The configurations are combined into the relaxation set and the \texttt{Pre-RP} is used to relax them. After the selection and DFT steps, the \texttt{Pre-RP} is re-trained. The cycle continues until the active learning convergence criteria is met, generating the \texttt{RP}. PRAPs contains several convergence criteria for active learning, which will be explained in Section \ref{sec:als}. If Step 1 was skipped, this training begins with an untrained MTP and an empty training set.

\textbf{Step 3:} The \texttt{AP} is trained next, also using active learning. While the \texttt{RP} is trained on the entire dataset, and thus is capable of predicting the EFS of any configuration in the system, the \texttt{AP} is designed to only be used on low-energy configurations. To start, the \texttt{RP} is used to relax both the \textbf{DFT\_CFG} and \textbf{URX\_CFG}. All configurations within 50~meV/atom of the lowest energy configuration for a particular composition are set aside into a new file: the low Energy Robust Relaxed (\textbf{lowE-RR}) set, which is used as the relaxation set for this step. The value of 50 meV/atom is a default and can be changed in the input file, described in Section~\ref{sec:input}.  The \texttt{Pre-RP} is also used as the initial MTP, but the training set for the \texttt{AP} starts off empty. 

\textbf{Step 4:} The final step is to perform analysis. Much of this step is optional, and full details are provided in Section \ref{subsec:chulls}. The \texttt{AP} is used to construct two new sets by both predicting and relaxing the \textbf{lowE-RR} set. Convex hulls for binary and ternary systems are generated from the initial \textbf{DFT\_CFG}, the \textbf{lowE-RR}, and the two sets obtained from the \texttt{AP}. If desired, PRAPs will also call AFLOW to relax every distinct structure via DFT to compare the MTP results against DFT. This step, however, can be computationally demanding as it requires DFT optimizations for a large number of structures. For higher-order systems, PRAPs may be able to list structures on, near, and above the convex hulls, but no visual plots are currently available. 

\subsection{Active Learning Scheme (ALS) Convergence} \label{sec:als}

The active learning loop needs to know when to stop. The default behavior follows Shapeev's recommendation: the ALS stops when no more configurations are selected to be placed in the training set~\cite{MLIP2}. When this happens, it is assumed that the MTP-based simulation could not produce any configurations sufficiently different from those in the training set to be worth adding (and if any were added, the MTP's predictions would not likely improve). This can take a long time, especially if the MLIP training and relaxation steps are running on a serial architecture. 

To reduce this computational expense, and hasten the time required to train the \texttt{AP}, especially for MTPs with a large MTP level (e.g.\ lev$_\text{max}\ge$~20), PRAPs contains other options that may end the ALS early at the cost of predictive accuracy. (i) The first option stops when the number of configurations to be added to the training set are less-than 1\% of the training set size. (ii) The second option will stop the ALS when the energy training error (measured as the root-mean-square error) is less-than a specific value (default of 0.05~eV/atom). (iii) The third option stops when the energy training error for the most recent step has changed by less-than 0.1~meV/atom from the previous step. (iv) The fourth and final option stops after 50 steps. Note that the default behavior is always checked for and takes priority, meaning that even if a user specifies a non-default convergence criteria, the default may be faster and the code will stop at that point.

\subsection{Measures of Performance}

PRAPs contains a few different ways to measure an MTP's predictive capability. The first is the error determination present in the MLIP package, which compares an MTP's predicted EFS against the EFS present in a particular .cfg file (typically DFT data).  Of this, two pieces of information are of particular interest: the mean-absolute-error (MAE) and the root-mean-square-error (RMSE) of the energy given in units of meV/atom. Training errors (calculated against the training set) are obtained for each trained MTP. Prediction errors (calculated against configurations not necessarily used in training) are obtained for various potentials. The \texttt{Pre-RP}, \texttt{RP}, and \texttt{AP} are all used to predict the EFS of the structures in the \textbf{DFT\_CFG} set, and the \texttt{AP} is also used to predict the EFS of the structures in the \textbf{lowE-RR} set. Most errors are simply recorded by PRAPs for examination by the user, but PRAPs does employ the training RMSE of the five pre-trained potentials to select the \texttt{Pre-RP}. 

PRAPs will index all of the configurations in \textbf{DFT\_CFG} and rank them in energy, high-to-low, then write the indices of the ten highest and the ten lowest energy configurations to file. After the pre-training step is complete, the five developed potentials are employed to predict the energies of the configurations within the \textbf{DFT\_CFG} set to determine which one of the five potentials performs the best in identifying the ten most and least stable configurations, and this metric is employed to choose the \texttt{Pre-RP} (see Section \ref{sec:workflow}). This same procedure is applied to the \textbf{lowE-RR} set, where the predictions of the high and low energy systems by the \texttt{RP} and \texttt{AP} are compared. The number of matching configurations is written to file for the user to inspect, but is not employed further.

\subsection{Filtration of Configurations} \label{sec:filt}

Key for the generation of reliable ML models is the curation of the training data, to remove outliers, unreliable data points, and to ensure that the structures comprising the data set are those that the model is intended for.  A few options are available in PRAPs to remove undesirable structures from a data set. First, PRAPs performs a distance-based filtration on all configurations before beginning any training. Any structures whose minimum interatomic distances are too large or too small (default 1.1 $\AA$ to 3.1 $\AA$) are removed. A user can also specify a volume-based or force-based filtration, in which configurations are kept only if their volume is within a particular range or if they possess no force components that are larger than a user-defined threshold. 

Two other forms of filtration are provided, though PRAPs does not use them automatically, as they may remove desirable structures. The first is trajectory filtering. Should a user have entire relaxation trajectories in their data, there is a means to select only the final-relaxed-structures. This was used in our previous paper~\cite{Zurek:2023n} to provide a standard set of structures against which to make predictions. The second is energy filtering. A user may specify a set, center, and spread for energies to be kept. For example, keep all structures within two standard deviations of the mean, or keep all structures within 50 meV/atom of the lowest energy structure for that composition. Full details are provided in the official manual attached as Supporting Information.

\subsection{Convex Hull Plots} \label{subsec:chulls}

For a binary or ternary system, PRAPs automatically generates a variety of convex hull diagrams. To calculate convex hulls, the energies of the elemental endpoints are required. We recommend the user provide a .cfg file with the ground-state elemental configurations (\textbf{REF\_CFG}). For convex hulls obtained by prediction or relaxation via MTP, the \textbf{REF\_CFG} will be obtained with the same procedure to ensure that the enthalpies of formation are computed using data calculated with the same level of theory.  For convex hulls obtained by DFT, the \textbf{REF\_CFG} will be relaxed with DFT.  If this file is not provided, an internal library of reference values will be employed. The values in this library were taken from AFLOW's online convex hull visualization tool, and should be treated with caution. 

The energies of the elemental phases are used for the calculation of the enthalpy of formation of the multicomponent system via $\Delta H_\text{F}=E^{\text{cfg}}-\sum_z E_z^{\text{ref}}n_z$,
where $E^\text{cfg}$ is the energy-per-atom of the configuration, predicted by the MTP or DFT, $E_i^\text{ref}$ is the reference energy-per-atom of each element $z$, and $n_z$ is the number of atoms of each element $z$ in the configuration. 

\begin{figure*}
    \centering
    \includegraphics[width=1.5\columnwidth]{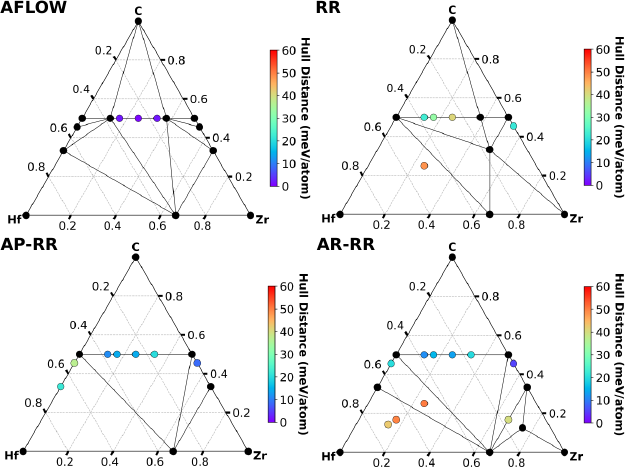}
    \caption{Examples of convex hull plots generated by PRAPs at Level 10. (a) The AFLOW-derived convex hull. (b) Relaxation of the AFLOW set with the Robust Potential yields the \textbf{RR} hull. (c) Prediction of the enthalpies of the \textbf{RR} structures with the Accurate Potential results in the \textbf{AP-RR} set, (d) while relaxation of the \textbf{RR} structures with the Accurate Potential yields the \textbf{AR-RR} set. Structures are colored (see color bar) according to the distances from the hull. Black dots are on the hull and purple dots are 1~meV/atom from the hull. This figure is adapted from material created by Josiah Roberts and provided in the Supporting Information of Ref.\ \cite{Zurek:2023n} (https://doi.org/10.1038/s41524-024-01321-7) licensed under a Creative Commons Attribution 4.0 International License (http://creativecommons.org/licenses/by/4.0/).}
   \label{fig:chull}
\end{figure*}

PRAPs will always produce the following five convex hulls: (i) one made of the filtered \textbf{DFT\_CFG}, (ii) one from the \textbf{lowE-RR}, (iii) one from the \texttt{AP}'s prediction of \textbf{DFT\_CFG} (\textbf{AP-v}), (iv) one from the \texttt{AP}'s prediction of the \textbf{lowE-RR} (\textbf{AP-RR}), and (v) one from the \texttt{AP}'s relaxation of the \textbf{lowE-RR} (\textbf{AR-RR}). The \textbf{AR-RR} is the `most processed' and represents the total action of MTPs on the original data and is, for many users, the final result. This procedure is also likely to provide the best predictions for other low energy compounds. For example, we recently showed that by first relaxing structures with the \textbf{RP} and subsequently with the \textbf{AP}, excellent predictive capability was obtained for the energies of an ensemble of phases that could be described as colorings of the hexagonal CMo/CW prototypes~\cite{Zurek:2023n}. Comparison of the convex hull obtained using the original \textbf{DFT\_CFG} data with the \textbf{AP-v} helps to determine if the \texttt{AP} can correctly predict the energies of structures comprising this set. Inspection of the \textbf{AP-RR} and \textbf{AR-RR} convex hulls reveals the effect of relaxation via the \texttt{AP} on the energies, as it can be substantial~\cite{Zurek:2023n}. PRAPs will plot all configurations on the convex hull in black, while those within a value of \textsl{Chull\_var} eV/atom (Table 1) above the hull are plotted using a gradient color scale. Configurations lying farther above the hull, and those with $\Delta H_\text{F} > 0$, are not plotted. Example plots are provided in Figure \ref{fig:chull}.

The plots described above are named `convex hull candidates' because the energies were predicted with MTPs. Further relaxation with DFT may change the geometrical parameters and energies of these configurations, leading to a new set of convex hull plots. This DFT relaxation can be optionally performed, and requested in the input file (Section \ref{sec:input}). PRAPs takes the five data sets used to plot the hull candidates, converts the configurations to POSCAR format, invokes AFLOW's prototype labels to find duplicate structures within each set, removes the duplicates, and submits the remaining configurations for DFT relaxation either via AFLOW or a user specified scheme. Because of the large number of DFT relaxations, this step is the most time and resource consuming out of the entirety of PRAPs. 

Once each of the five sets, \textbf{lowE-RR}, \textbf{AP-v}, \textbf{AP-RR} and \textbf{AR-RR}, are relaxed with DFT, they are concatenated with the DFT relaxed structures in \textbf{DFT\_CFG}, resulting in a further set of plots. This final set ensures that the structures that appeared during the PRAPs procedure really are on the convex hull, when compared against the original set of structures used for training the MTP. For example, in our previous study of ternary carbides, this step was employed to determine if any of the PRAPs found structures were on the hull when the data in AFLOW was taken into consideration~\cite{Zurek:2023n}. After calculation, PRAPs generates image files and a few summary files. PRAPs also contains optional convex hull diagrams that show hull distance and enthalpy for all configurations, even those far above the convex hull.

\begin{table*}[h!]
    \centering
    \begin{tabular}{|c|c|c|l|}
        \hline 
        \textbf{Functions} & \textbf{Description} & \textbf{Required}? \\
        \hline \hline
    \textsl{els} & The elements given as a Bash array. & Req. \\
    \hline
    \textsl{Lev\_MTP} & The level of MTP used for training. & Req. \\
    \hline
    \textbf{DFT\_CFG} & The .cfg file with EFS information. & Opt.$^a$ \\
    \hline
    \textbf{URX\_CFG} & A .cfg file without EFS information. & Opt.$^a$ \\
     \hline
    \textbf{REF\_CFG} & A .cfg file with ground-state elements. & Opt. \\
    \hline
    \textsl{cmpd\_pth} & The working directory, defaults to \emph{install\_directory/els} & Req. \\
     \hline
    \textsl{Chull\_var} & Defines the energy range of the \texttt{AP}, default 0.05~eV/atom. & Opt. \\
     \hline
    \textsl{mindist} & The smallest acceptable minimum interatomic distance. Default 1.1 $\AA$ & Opt.\\
     \hline
    \textsl{maxdist} & The largest acceptable minimum interatomic distance. Default 3.1 $\AA$ & Opt. \\
     \hline
    \textsl{relax\_settings} & Desired settings for MLIP's relaxation command. & Opt. \\
     \hline
    \textsl{training\_settings} & Desired settings for MLIP's training command. & Opt. \\
     \hline
     \textsl{basic\_acc} & Trains an additional \texttt{AP} with basic training. Defaults to False. & Opt. \\
     \hline
    \textsl{CHULL} & Whether or not to calculate all convex hulls. Defaults to False. & Opt. \\
     \hline
     \textsl{save\_outcars} & Saves all OUTCAR files from VASP calculations. Defaults to False. & Opt. \\
     \hline
    \textsl{custom\_relax} & If \textsl{CHULL} = True, this replaces AFLOW's DFT with your own. & Opt. \\
     \hline
    \textsl{filter\_trajectories} & Selects the final relaxed structures for \texttt{AP} training. Defaults to False. & Opt. \\
     \hline
     \textsl{filter\_volumes} & Filters the \textbf{DFT\_CFG} by volume instead of \textsl{mindist}. Defaults to False. & Opt. \\
     \hline
     \textsl{volume\_scaling} & Keep all structures $1-x < V < 1+x$, where $x$ is a decimal from 0 to 1. & Opt. \\
     \hline
     \textsl{filter\_forces} & Filters the \textbf{DFT\_CFG} by forces. Defaults to 5.0, set to 0 to disable. & Opt. \\
     \hline
     \textsl{ALS\_conv} & The convergence criteria 0-4 as described in Section 3.2. Default 0. & Opt. \\
     \hline
     \textsl{CHK} & Checkpoint tag 0-4, see text for explanation. & Opt. \\
     \hline
    \end{tabular} \\
    $^a$ At least one of these must be present.
    \caption{Summary of tags in the input file. Some tags are required (Req.) while others are optional (Opt.). A full description of these can be found in the manual (see the Supporting Information).}
    \label{table:inpraps}
\end{table*}

\subsection{Installation and Runtime}

In this section we list the programs that are required for PRAPs to run, and basic installation instructions, which are expanded upon in Section 1 of the Supplementary Information.
PRAPs requires  the following software:  VASP~\cite{VASP},  MLIP~\cite{MLIP2}, as well as the Pandas and Matplotlib Python modules. For the optional analysis steps (Step 3 in Figure \ref{fig:workflow}) AFLOW~\cite{AFLOW,AFLOWstd} is required. Upon downloading the PRAPs software, a user should read the README, unpack the tar archive, and run the installation script \emph{adjust\_paths.py}. Users must specify, as arguments, the locations of: the PRAPs install directory, the VASP POTCAR files, the untrained MTPs in the MLIP install directory, the user's Slurm module files, and the user's Python libraries. The installation will create a \emph{.../PRAPs/} directory in the designated place, containing \emph{ser/}, \emph{par/}, \emph{utils/}, and \emph{examples} subdirectories for serial MLIP, parallel MLIP, utilities, and tutorials, respectively.  The main contents will be scripts in Bash and Python, along with example Slurm submission scripts, and an example PRAPs input file: \emph{inpraps.sh}.  

When running PRAPs, a user can look for the example Slurm script, which issues the following command:
{\begin{lstlisting}[language=bash]
bash $PRAPs_PATH/par/praps.sh inpraps.sh
\end{lstlisting}}
\noindent A user may point the script to the correct install location, the correct serial or parallel version, and the correct \emph{inpraps.sh}, as desired. Note that PRAPs expects \emph{inpraps.sh} and all other input .cfg files to be in the same location as the output will be written. Users may need to adjust this Slurm script for their particular cluster, or to adapt job submission scripts that are compatible with the cluster management and job scheduling system that works on their computing cluster. PRAPs should still run fine, as the primary calls to execution are Bash and Python.

Unlike VASP, PRAPs is intended to be run from a central directory, defaulting to the install location. When run, PRAPs will note the submission directory and make a sub-directory for the elemental system specified, such as CHfTa, CrMn, or AgAu (to ensure compliance with AFLOW, PRAPs records elements in alphabetical order). All of the output will be placed in this elemental sub-directory. If a user desires different behavior, they may specify where to write data in the input file. 

To run PRAPs, a user needs one or both of the \textbf{DFT\_CFG} and \textbf{URX\_CFG} configuration files. The first contains EFS data obtained from DFT, and the second only contains structural coordinates. If a user does not provide the \textbf{DFT\_CFG} file, they must also indicate to skip the pre-training step in \emph{inpraps.sh}. The \emph{inpraps.sh} file contains many of the MLIP and PRAPs-specific settings for the run, described fully in Section \ref{sec:input}. The most important are the elements, in the correct order, the level of MTP, the filenames of the \textbf{DFT\_CFG} and/or the \textbf{URX\_CFG} files, and the path where the output should be directed. 

When the process is finished, PRAPs will clean up by creating a .tar archive containing most of the files generated during the process, especially the MTPs, error files, and the various inputs. All remaining temporary files will be deleted. A copy of the MTPs will be placed in a directory above the working directory; this directory is named \emph{pots}.

\section{Input File} \label{sec:input}

Below we provide an example of the input file used to run PRAPs, given as a bash script, followed by an explanation of the keywords along with their accepted values in Table \ref{table:inpraps}.
{\begin{lstlisting}[language=bash]
PRAPs Input File -- inpraps.sh
els = (C Hf Ta)
LevMTP = 10
DFT_CFG = example.cfg
URX_CFG = example_2.cfg
REF_CFG = example_3.cfg
cmpd_pth = ./special_path/
Chull_var = 0.05
mindist = 1.1
maxdist = 3.1
relax_settings = "--limit=100"
training_settings = "--name=post.mtp"
basic_acc = false
CHULL = false
save_outcars = false
custom_relax = false
filter_trajectories = false
filter_volumes = false
filter_forces = 0
volume_scaling = 0.25
ALS_conv = 0
CHK = 0
\end{lstlisting}}

In addition, we briefly explain the behavior of  the optional \textsl{CHK} or checkpoint tag, whose default value is 0. This tag should be omitted for a completely new PRAPs run. Over time, \textsl{CHK} tags are appended into \emph{inpraps.sh}. PRAPs looks for the largest value of \textsl{CHK},  and skips all steps beforehand. The value of this tag can range from 0-1, and it signifies the step in Figure \ref{fig:workflow} after which the workflow will begin. The default of 0 starts at the pre-training step, while setting \textsl{CHK} = 1 avoids the pre-training step, and in the event of a crash or interruption (such as exceeding a wall-time-limit) values of 2, 3 or 4 will restart the job at roughly the same point it was stopped. 

\subsection{Additional Utilities} \label{sec:utils}

PRAPs contains a few other utility functions, for which a very brief description follows. Please see the manual in the Supplementary Information for the full details. PRAPs comes with a Python library titled \textit{mliputils}. This library performs functions such as reading, writing, and filtering the .cfg file by converting into a data table. One notable operation is the conversion of POSCAR files to .cfg format, a utility not provided by the MLIP software. Many of the mathematical and filtration operations are performed using this library. Users may filter .cfg files by energy, force, minimum interatomic distance, and a few other criteria. A brief overview of functions is provided in Table \ref{table:utils}; for more detailed instructions please refer to the official manual. There is also a tracking system users may use (PRAPs-ID). This ID system helps users keep track of changes made to specific configurations during the PRAPs process. While much of the implementation is handled automatically, documentation exists for users who wish to customize their experience or perform tasks with MLIP that are not already handled by PRAPs.

\begin{table*}[ht!]
    \centering
    \begin{tabular}{|c|l|}
        \hline 
        \textbf{Functions} & \textbf{Description} \\
        \hline \hline
                                                            & Reads a .cfg file, and returns a DataFrame (cfg-df). \\        
    \textsl{read\_cfg\_from\_file}           & \textbf{Arguments:}                                   \\
                                                             & filename - string or path                             \\
    \hline
     & Reads a .json generated from AFLOW and returns a DataFrame (cfg-df). \\
    \textsl{read\_json} & \textbf{Arguments:} \\
     & filename - string or path \\
     \hline
     & Given a structure in POSCAR file format, adds a new entry in the cfg-df. \\
    \textsl{read\_cfg\_from\_poscar} & \textbf{Arguments:} \\
     & cfg - new or working DataFrame \\
     & els - list of elements in type-order \\
    \hline
     & Writes a cfg-df to a .cfg file. \\
    \textsl{write\_cfg} & \textbf{Arguments:} \\
     & cfg - the working cfg-df \\
     & filename - string or path \\
     & mode - a (append) or w (write, default) \\
     & start - DataFrame index to start writing \\
     & stop - DataFrame index to stop writing \\
     \hline
     & Returns a new cfg-df with configurations below the specified energy limit. \\
    \textsl{get\_low\_E} & \textbf{Arguments:} \\
     & cfg - the working cfg-df \\
     & lim - a float, the maximum desired energy in eV/atom (default 0.05) \\
     \hline
     & Adds a composition column to the cfg-df. \\
    \textsl{get\_comp} & \textbf{Arguments:} \\
     & cfg - working cfg-df \\
     & style - 1, 2, 3 (see manual) \\
     & typedict - type-element dictionary \\
     \hline
     & Filters configurations by energy. \\
    \textsl{clean\_df} & \textbf{Arguments:} \\
     & cfg - working cfg-df \\
     & method - string (see manual) \\
     \hline
     & Returns elemental ground state energies if present in the cfg-df. \\
    \textsl{get\_min\_endpoints\_from\_cfg} & \textbf{Arguments:} \\
     & cfg - working cfg-df \\
     \hline
     & Returns elemental ground state energies using internal dictionary. \\
    \textsl{get\_min\_endpoints\_from\_els} & \textbf{Arguments:} \\
     & els - list of elements \\
     \hline
     & Adds a column for enthalpy of formation to the cfg-df. \\
    \textsl{get\_Hf} & \textbf{Arguments:} \\
     & cfg - working cfg-df \\
     & endpts - dictionary of elemental energies \\
     \hline
     & Generates a convex hull using Scipy. \\
    \textsl{convexhull} & \textbf{Arguments:} \\
     & cfg - working cfg-df \\
     \hline
     & Calculates hull distances for configurations above the hull. \\
    \textsl{chull\_dist} & \textbf{Arguments:} \\
     & hull - Scipy convex hull object \\
     & points - Scipy convex hull points object \\
     \hline
    \end{tabular}
    \caption{Summary of useful functions in the mliputils library. A full description of these, and other, less useful functions, can be found in the manual (see the Supporting Information).}
    \label{table:utils}
\end{table*}

\section{Practicalities}

\subsection{Sample Data}

\begin{figure*}
    \centering
    \includegraphics[width=1.5\columnwidth]{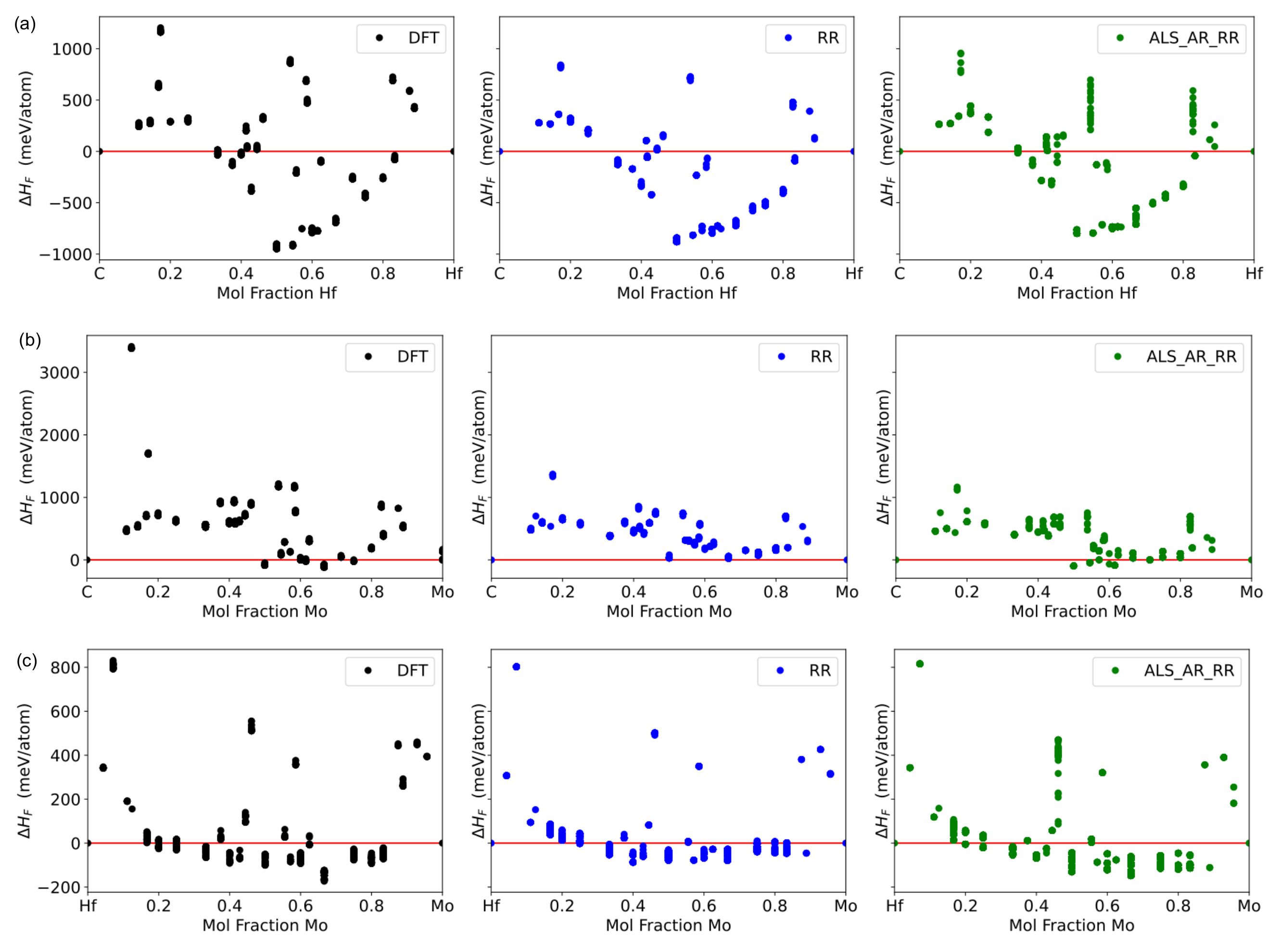}
    \caption{Examples of 2D convex hull plots generated by PRAPs for the (a) CHf, (b) CMo, and (c) HfMo systems.  The AFLOW-derived convex hulls are on the left. Subsequent relaxation of those structures by the Robust Potential gives the middle plots. A second relaxation by the Accurate Potential yields the plots on the right.}
   \label{fig:chfmo}
\end{figure*}

We provide a few sets of sample data for users to use in testing for the following binary systems: CHf, CMo, and HfMo. This data comprises configurations and corresponding DFT data from the AFLOW~\cite{AFLOW} relaxation trajectories (10,231, 13,551 and 12,807 configurations, respectively) along with 111 cubic and hexagonal structures generated by \textsc{RandSPG}~\cite{RandSPG}. These tests can be performed by running PRAPs at Level 16 using the supplied files in the PRAPs package. The errors obtained for the \texttt{RP} and \texttt{AP} obtained with PRAPs are shown in  Table \ref{table:errs}, and the convex hulls appear in Figure \ref{fig:chfmo}. It should be noted that PRAPs provides a number of options for convex hull plots, only one of which is illustrated here. Because MLIP initializes each MTP with random parameters, and the configurations used to train the MTP are chosen randomly, the results obtained from different PRAPs run will not be identical. 

\begin{table*}[ht!]
    \centering
    \begin{tabular}{|c|l|}
        \hline 
        \textbf{System} & \textbf{Training Error (meV/atom)} \\
        \hline \hline
     CHf-16 Pre-RP & 62 (89) \\
     \hline
     CHf-16 RP & 115 (181) \\
     \hline
     CHf-16 AP & 26 (45)  \\
     \hline
     CMo-16 Pre-RP & 103 (191) \\
     \hline
     CMo-16 RP & 125 (217) \\
     \hline
     CMo-16 AP & 21 (30) \\
     \hline
     HfMo-16 Pre-RP & 31 (87) \\
     \hline
     HfMo-16 RP & 45 (93) \\
     \hline 
     HfMo-16 AP & 12 (18) \\
     \hline
    \end{tabular}
    \caption{Summary of training errors for the sample data given as mean-absolute-error and root-mean-squared-error (in parentheses).}
    \label{table:errs}
\end{table*}

\subsection{Universal Potentials}

While ML-IAPs (such as the MTP models used in this study) have traditionally been created for specific chemical systems, recent advances have led to the development of universal interatomic potentials (UIPs) designed to provide a data-driven description of interatomic interactions over a wide range of elements \cite{MACE,chgnet,rhodes2025orbv3,mattersim,m3gnet, alignn,GoogUniverse}. UIPs have demonstrated promising performance in various applications; however, their degree of transferability remains a subject of ongoing research~\cite{MACE}. Given the growing popularity and potential impact of UIPs, we have conducted a comparison of convex hulls generated by select UIPs with our own \texttt{RP} and \texttt{AP}. The latest release of \textsc{XtalOpt}~\cite{XtalOpt14} includes an interface script to perform local optimizations using a variety of UIPs. For this benchmarking test, we utilized this interface script to perform structural relaxations with the commonly-used MACE~\cite{MACE} and MatterSim~\cite{mattersim} potentials, and re-produced the corresponding convex hulls.

\begin{figure*}
    \centering
    \includegraphics[width=1.5\columnwidth]{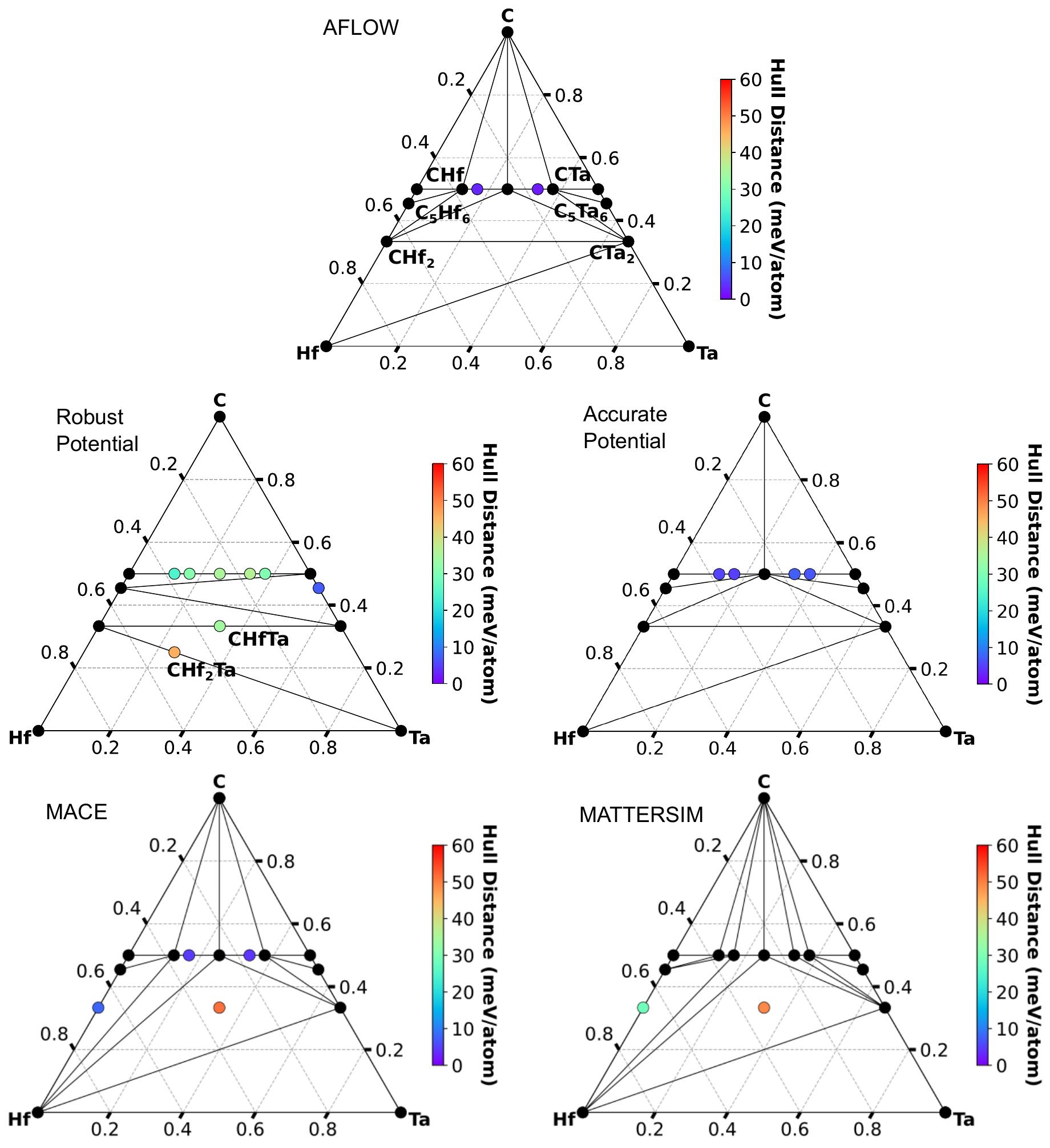}
    \caption{Examples of convex hull plots generated by PRAPs compared with universal interatomic potentials (UIPs). Structures displayed are within 60~meV/atom of each hull and are colored (see color bars) according to distances from the hull. (a) The AFLOW-derived convex hull. The middle row shows hulls derived by relaxing structures with: (b) the robust potential, and (c) subsequent relaxation with the accurate potential. The bottom row shows convex hulls derived, from the same data, from UIPs without any action by an MTP, showing (d) MACE and (e) MatterSim. Black dots are on the hull and purple dots are 1~meV/atom from the hull. The data has not undergone further DFT relaxations. Portions of this figure are adapted from material created by Josiah Roberts and provided in the Supporting Information of Ref.\ \cite{Zurek:2023n} (https://doi.org/10.1038/s41524-024-01321-7) licensed under a Creative Commons Attribution 4.0 International License (http://creativecommons.org/licenses/by/4.0/).}
   \label{fig:uip}
\end{figure*}

To make this comparison we relaxed the $\sim$12k structures for the CHfTa system from our previous publication~\cite{Zurek:2023n}, which consisted of configurations taken from relaxation trajectories within AFLOW~\cite{AFLOW} along with cubic and hexagonal structures generated by \textsc{RandSpg}~\cite{RandSPG}, with the considered UIPs. Duplicates were removed post relaxation. Those structures that were within 60~meV/atom of the convex hull are plotted in Figure \ref{fig:uip}, which also shows the original AFLOW data and the hulls generated by the \texttt{RP} and \texttt{AP} at an MTP Level of 16 for comparison.  MACE and MatterSim yield convex hulls that appear to have a slightly better agreement with the AFLOW data as compared to the \texttt{RP} generated convex hull. Nonetheless, all three models predict a low-energy CHfTa phase that is not present within 60~meV/atom of the AFLOW or \texttt{AP} hulls. These results suggest that the recently released UIPs could be used to make a preselection of the low-energy structures employed for the generation of the \texttt{AP}, forgoing the training of the \texttt{RP} entirely, but the tailored \texttt{AP} improves the predictions for those structures near the hull as compared to the UIPs. At the moment, this type of workflow is not automated within PRAPs. We anticipate better integration of UIPs into PRAPs so that users may select their preferred relaxation tool, in a future update, which may also include extensions to other DFT codes, such as Quantum Espresso.

\subsection{Computational Cost}

In the PRAPs workflow the DFT calculations determine the computational overhead; the MTP training and relaxation, and the functions of PRAPs are comparatively fast, taking on the order of minutes. Our initial paper, which applied PRAPs to ternary carbides~\cite{Zurek:2023n}, introduced the CHfTa systems appearing in Figure \ref{fig:uip}. For this system $\sim$12,000 configurations were locally optimized with PRAPs, creating both an \texttt{AP} and an \texttt{RP}. The number of calls for DFT single-point calculations performed during the active learning procedure  to generate the \texttt{AP} was about double the number required for the \texttt{RP} for this specific case. While the \texttt{AP} will always have a higher computational overhead than the \texttt{RP}, the exact ratio is likely system and convergence dependent. Futhermore, higher MTP levels, which have a larger number of adjustable parameters, required a larger number of single-point DFT calculations. Based upon these tests, we concluded that an MTP level of 16 presented a good balance between accuracy and computational expense, requiring around 1000 DFT calls for the CHfTa system.

Once potential training is complete PRAPs can, optionally, generate a number of convex hulls for analysis. In this step PRAPs uses AFLOW to assist in determining and removing duplicate structures, but a final DFT relaxation of all unique structures is performed. As noted in Section \ref{subsec:chulls}, this step requires the most time and resources of the whole PRAPs procedure.

\section{Conclusions}
The PRAPs package is an integrated workflow for training moment tensor potentials and provides convenient convex hull analysis. The package also comes with a variety of functional utilities not present in the MLIP software package, including conversion of POSCAR to .cfg file formats, integration with AFLOW, manipulation of data in the .cfg file, generation of certain plots of interest, and an ID system to track changes to individual structures in the .cfg file over time. Future efforts and updates will focus on integration with Quantum Espresso, improvements with LAMMPS, and a Monte Carlo algorithm for structure prediction that employs energies computed with MTPs. \\

\noindent\textbf{CRediT authorship contribution statement} 

\textbf{Josiah Roberts}: Writing, Software, Investigation, Methodology, Data Curation;
\textbf{Biswas Rijal}: Writing \& Editing, Validation;
\textbf{Simon Divilov}: Validation, Resources;
\textbf{Jon-Paul Maria}: Conceptualization, Funding Acquisition;
\textbf{William G. Fahrenholtz}: Conceptualization, Funding Acquisition;
\textbf{Douglas E. Wolfe}: Conceptualization, Funding Acquisition;
\textbf{Donald W. Brenner}: Conceptualization, Funding Acquisition;
\textbf{Stefano Curtarolo}: Conceptualization, Project Administration, Funding Acquisition;
\textbf{Eva Zurek}: Writing - Review \& Editing, Supervision, Software, Resources, Project Administration, Funding Acquisition, Conceptualization.
 \\[1ex]

\noindent\textbf{Declaration of competing interest}
The authors declare that they have no known competing financial interests or personal relationships that could have appeared to influence the work reported in this paper. \\[1ex]

\noindent\textbf{Data availability}
No data was used for the research described in the article. \\[1ex]

\noindent\textbf{Acknowledgement}
We would like to gratefully acknowledge the DoD SPICES MURI sponsored by the Office of Naval Research (Naval Research contract N00014-21-1-2515) for financial support, the Center for Computational Research at SUNY Buffalo (http://hdl.handle.net/10477/79221) for computational support, and Samad Hajinazar for fruitful discussions.\\[1ex]


\bibliographystyle{elsarticle-num}
\bibliography{refs}

\end{document}